\documentclass[12pt,preprint]{emulateapj}
\slugcomment{{\sc Accepted to ApJL:} February 28, 2013} 

\shortauthors{YAMADA ET AL.}
\shorttitle{RAPID SPECTRAL CHANGE IN HARD X-RAY OF CYGNUS X-1 WITH SUZAKU}

\begin{document} 
\title{RAPID SPECTRAL CHANGES OF CYGNUS X-1 IN THE LOW/HARD STATE WITH SUZAKU} 


\author{S. Yamada\altaffilmark{1}, 
H. Negoro\altaffilmark{2}, 
S. Torii\altaffilmark{3}, 
H. Noda\altaffilmark{3}, 
S. Mineshige\altaffilmark{5}, 
and K. Makishima\altaffilmark{3,1}
}

\altaffiltext{1}{Cosmic Radiation Laboratory,
 Institute of Physical and Chemical Research (RIKEN),
   Wako, Saitama, 351-0198, Japan}
\altaffiltext{2}{Department of Physics, College of Science and Technology, Nihon University, 1-8 Kanda-Surugadai, Chiyoda-ku, Tokyo 101-8308}   
\altaffiltext{3}{Department of Physics, University of Tokyo, 7-3-1 Hongo, Bunkyo-ku, Tokyo, 113-0033, Japan}
\altaffiltext{4}{Institute of Space and Astronautical Science, JAXA, 
3-1-1 Yoshinodai, Sagamiharas, Kanagawa, Japan 229-8510} 
\altaffiltext{5}{Department of Astronomy, Kyoto University, 
Kitashirakawa Oiwake-cho, Sakyo-ku, Kyoto 606-8502, Japan}
\altaffiltext{6}{Department of Electronic Information Systems, Shibaura Institute of Technology,
307 Fukasaku, Minuma-ku, Saitama-shi, Saitama, 337-8570, Japan} 

\begin{abstract}


Rapid spectral changes in the hard X-ray on a time scale down to $\sim0.1$ s 
are studied by applying ``shot analysis" technique 
to the {\it Suzaku} observations of the black hole binary Cygnus X-1, 
performed on 2008 April 18 during the low/hard state. 
We successfully obtained the shot profiles 
covering 10--200 keV with the Suzaku HXD-PIN and HXD-GSO detector. 
It is notable that the 100-200 keV shot profile is acquired for the first time 
owing to the HXD-GSO detector. 
The intensity changes in a time-symmetric way, 
though the hardness does in a time-asymmetric way.  
When the shot-phase-resolved spectra are quantified with the Compton model, 
the Compton $y$-parameter and the electron temperature are
found to decrease gradually through the rising phase of the shot, 
while the optical depth appears to increase. 
All the parameters return to their time-averaged values 
immediately within 0.1 s past the shot peak. 
We have not only confirmed 
this feature previously found in energies below $\sim$~60 keV, 
but also found that the spectral change is more prominent in energies above $\sim$~100 keV, 
implying the existence of some instant mechanism for direct entropy production. 
We discuss possible interpretations on the rapid spectral changes in the hard X-ray band. 


\end{abstract}


\section{INTRODUCTION}

Starting with the first identification of the black hole (BH) binary Cygnus X-1 (hereafter Cyg
X-1) in the early 1970's (e.g., Oda et al.~1971; Tananbaum et al.~1972; Thorne and Price~1975), 
X-ray observations have been playing 
an important role to reveal spectral and temporal properties of BH binaries, 
which are largely classified into two distinct states: the high/soft state and the low/hard state 
(e.g., Remillard and McClintock 2006; Done et al.~2007). 
In contrast to the high/soft state characterized by the dominant disk emission 
(Mitsuda et al. 1984; Makishima et al, 1986) from the standard disk (Shakura \& Sunyaev 1973), 
the spectrum in the low/hard state is expressed by a powerlaw with a photon index of $\sim$ 1.5
with an exponential cutoff at $\sim$ 100 keV (e.g., Sunyaev \& Tr\"{u}mper 1979) 
from a hot ``corona'' (e.g., Ichimaru 1977; Narayan \& Yi 1995; 
chapter~8 in Kato, Fukue, and Mineshige 2008). 
Rapid time variabilities on a time scale of $\sim$ ms (e.g., Miyamoto et al. 1991) 
only seen in the low/hard state have been studied in many ways 
(e.g., Nowak et al. 1999; Poutanen 2001; Pottschmidt et al. 2003; Uttely et al.~2011; Torii et al. 2011), 
though the origin is still missing piece of puzzle,  
presumably due to observational difficulties of realizing 
both high sensitivity and large effective area. 

A distinctive approach is ``shot analysis'' (Negoro et al.~1994; Negoro~1995) 
adopted for Cyg X-1 obtained with {\it Ginga}. 
This method is the time-domain stacking analysis 
to obtain universal properties behind non-periodic variability. 
It is in time-domain analysis that 
we can combine spectral information in a straightforward way.   
They found three main features: 
(1) the intensity changes time symmetrically,  
(2) both of the rise and decay curves are well represented 
by the superpositions of two exponential functions 
with time constants of $\sim 0.1$ s and $\sim 1.0$ s, 
and (3) the spectral variation is, by contrast, time asymmetric in the sense 
that it  gradually softens toward the peak and instantly hardens across the peak 
(see Figure 2 in Negoro et al.~1994). 
These properties are further investigated with {\it RXTE} (Focke et al.~2005).
The time constant of $\sim$ 1 s far exceeds the local (dynamical or thermal) timescale of the innermost region,
and should thus reflect accreting motion of gas element.
Manmoto et al. (1996) proposed an interesting explanation that  
inward-forwarding accreting blob, causing an increase in X-ray flux, 
are reflected as sonic wave when it reaches the BH (Kato, Fukue, and Mineshige 2008), 
though further observational constraints have been awaited. 

The extension of this approach towards higher higher energies, $\sim$ 200 keV, 
should be crucial because it may provide a hint to 
the physics causing the rapid spectral variation. 
Thus, we observed Cyg X-1 in the low/hard state with {\it Suzaku} (Mitsuda et al.~2007),  
by utilizing both the XIS (Koyama et al. 2007) 
located on the focus of the X-ray mirror (Serlemitsos et al. 2006) 
and the Hard X-ray Detector (HXD: Takahashi et al. 2007;
Kokubun et al. 2007; Yamada et al. 2011)
(Takahashi et al. 2007; Kokubun et al. 2007; Yamada et al. 2011).
The distance, the mass, and the inclination of Cyg X-1 are 
1.86$^{+0.12}_{-0.11}$ kpc (Reid et al.~2011; Xiang et al.~2011), 
$14.8\pm1.0 M_{\odot}$, and $27.1\pm0.8^{\circ}$ (Orosz et al.~2011), respectively.  
It has an O9.7 Iab supergiant, HD 226868 (Gies \& Bolton 1986) 
with an orbital period of $5.599829$ days (Brocksopp et al. 1999). 
Unless otherwise stated, errors refer to 90\% confidence limits.

\begin{figure*}
\begin{center}
\vbox{}
\includegraphics[width=0.95\textwidth]{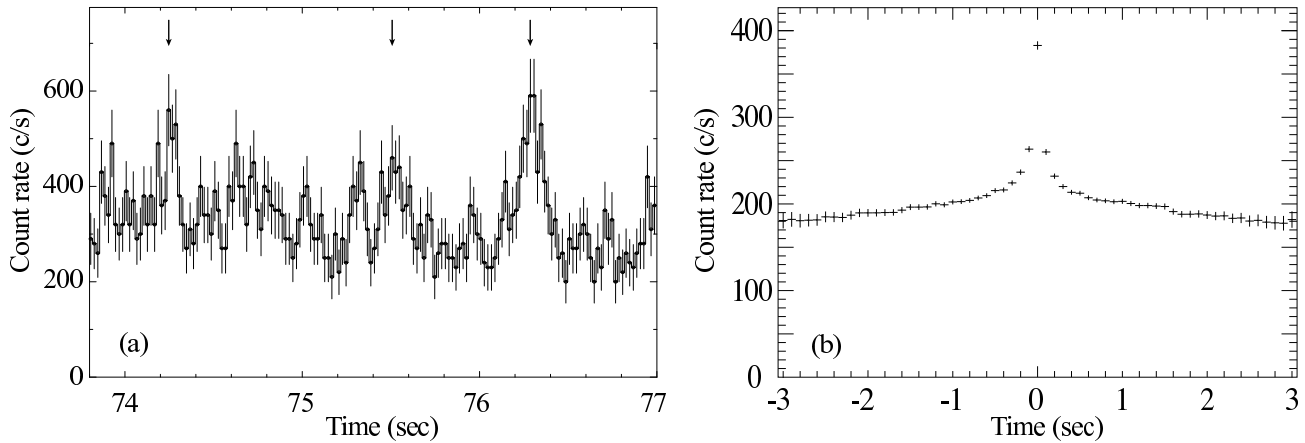}
\caption{
(a) A small portion for 3 s of a 0.5--10 keV light curve of Cyg X-1 taken with XIS0 operated in the P-sum mode. The time-bin size is 0.1 s. The arrows indicate the peaks judged as the shots. (b) The XIS0 shot profile created from all the light curve. }
\label{fig:lc}
\end{center}
\end{figure*}

\section{OBSERVATION AND DATA REDUCTION} 

Cyg X-1 data taken with {\it Suzaku} on 2008 April 18 
(ObsID=403065010) are used in this letter, 
which is one of the 25 observations in its low/hard state (see Yamada 2011 in details). 
The XIS0 was operated in the timing mode, or the Parallel-Sum (P-sum) mode, 
which is one of the clocking modes in the XIS. 
A timing resolution of the P-sum mode is $\sim$ 7.8 ms.  
Data reduction of the timing mode are different from the standard one 
in the Grade selection criteria\footnote{see http://www.astro.isas.ac.jp/suzaku/analysis/xis/ \\psum\_recipe/Psum-recipe-20100724.pdf}: Grade~0 (single event), Grade~1, and 2 (double events) are used.
The XIS background is not subtracted because it is less than 0.01~\% of the signal events. 


The HXD data consisting of the PIN~(10--60 keV) and GSO~(50--300 keV) events 
are processed in the same manner as Torii et al.~(2011). 
The events are selected by the criteria of elevation angle $\geq$ $5^{\circ}$, 
cutoff rigidity $\geq$ 6 GV, and 500 s after and 180 s before the South Atlantic Anomaly. 
The non X-ray background (NXB) of the HXD 
modeled by Fukazawa et al.~(2009) are subtracted from the HXD data. 
The NXB model reproduces the black-sky data with an accuracy of $\sim$ 1\% when the exposure is longer than $\sim$ 40 ks. 
The cosmic X-ray background can be ignored in our analysis, 
since its contribution is less than $\sim$ 0.1\%. 
The simultaneous exposure for the XIS0, PIN and GSO is 33.9~ks.

\section{DATA ANALYSIS AND RESULTS}

\subsection{Definition and preparation}

We formulate the definition of the ``shot''. 
Here $t$ and $C$($t$) is the event arriving time and the count rate at $t$.
$T$ is an interval of time over which $C$($t$) is averaged, 
and $\overline{C(t)}_T$ denotes the average count rate over an interval of $t - T <  t  < t + T$. 
$t_a$ and $t_b$ refer to any time after and before $t$,
satisfying the conditions of $t < t_a  < t + T$ and $ t - T < t_b < t $,
and $f$ is the dimensionless parameter of order unity for the threshold. 
The peak time $t_p$ of each ``shot'' is a local maximum in $C(t)$ defined as,   
\begin{equation}
\{  t_p  \mid  C(t_b)  < C(t) \geq C(t_a) \mathrm{~\&~} C(t) > f \overline{C(t)}_{T}  \}
\end{equation}
Equation (1) is nearly the same as that used in Negoro et al.~(1994) and Focke et al.~(2005), 
which works robustly since no iteration is incorporated in this process. 
The only caveat is that when accidentally more than two adjacent bins of $C(t)$ 
have the same value at the peak, the first of them is selected as the peak;  
e.g., when $C(t_1) < C(t_2) = C(t_3) > C(t_4)$ is realized, 
$t_2$ is to be the peak.  
It is possible to impose optionally another constraint on the separation of time 
between the two successive peaks $\Delta t$; 
e.g., $\Delta t  \geq g T$ works to avoid accumulating the small peaks 
or fake events caused by the Poisson statistics, 
where $g$ is the dimensionless parameter of order unity. 
Since we aimed at accumulating as many photons as possible 
to quantify spectral change, 
we adopted $g = 1$, which means that a minimum of $\Delta t$ equals $T$. 

An example of the light curve segment of XIS0 is shown in Figure~1(a). 
The events with 7.8 ms time resolution were binned into 0.1 s bins, 
resulting in 20--60 counts per bin.  
Intensity changes by a factor of $\sim$ 2, 
so that Poisson fluctuation ($\sim$ 15\% at 1$\sigma$) 
is sufficiently smaller than the intrinsic variation. 
Employing $f=1.0$ and $T=1.5$ s, 
we actually applied the procedure of Equation (1) 
to the entire P-sum light curve, 
and identified 7524 shots in total. 
The distribution of $\Delta t$ becomes 
a grossly exponential distribution 
in agreement with the previous reports (Focke et al.~2005). 
Figure~1(b) shows the shot profile obtained by stacking the 7524 shots 
with reference to each peak. 
We can see two exponential slopes 
in the shot profile; shorter and longer decay time constants of $\sim0.1$ s 
and 1--2 s.
Our primary focus is to extend this analysis to the HXD band, 
so we do not further investigate the shot profile 
and the XIS0 spectra due to incomplete calibration of the P-sum mode. 

\subsection{Shot profiles of the HXD data}  

\begin{figure}[b]
\begin{center}
\vbox{}
\includegraphics[width=0.46\textwidth]{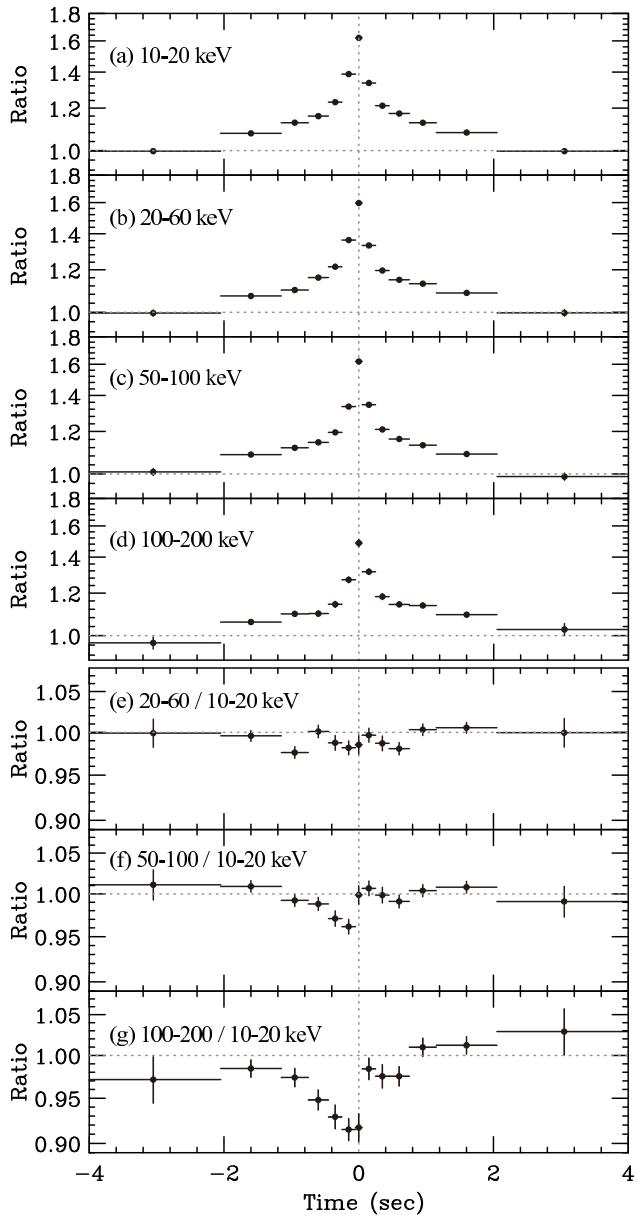}
\caption{
The stacked profiles of the shots in the HXD bands by using the XIS0 light curve as a reference. 
(a)--(d) The background-subtracted and deadtime-corrected shot profiles 
in 10--20 keV, 20--60 keV, 50--100 keV, and 100--200 keV, 
which are renormalized by the individual averages over -4 to -2 and 2 to 4 s. 
(e)--(f) The hardness ratios of the shot profiles. 
The profiles in (b)--(d) are divided by the 10--20 keV profile in (a).}
\label{fig:avespec} 
\end{center} 
\end{figure} 


According to the peak time determined with the XIS0 light curve, 
we have accumulated the PIN and GSO events and their NXB events\footnote{http://heasarc.gsfc.nasa.gov/docs/suzaku/analysis/abc/}.  
To estimate pileup effects in a phenomenological way, 
we purposely tried to use either 1${\arcmin}$ inside or outside image core of XIS0 
to obtain the shot profiles based on Yamada et al. (2012), 
though the shot profiles were not significantly changed. 
To investigate the energy dependence of the shots, 
we have utilized four energy bands:  
10--20 keV and 20--60 keV from PIN, and 50--100 keV and 100--200 keV from GSO. 
After subtracting the NXB events from the data and correcting them for dead time, 
we have obtained the 10--200 keV shot profiles with the HXD data. 
The stacked shot profiles 
are divided by the count rates averaged over -4 to -2 s and 2 to 4 s 
to approximately correct them for the differences in the efficiency. 

The normalized shot profiles are shown in Figure~2(a)--(d). 
The derived profiles appear all very similar in shape to the one of XIS0. 
However, energy dependencies are certainly found when their widths of the peaks are carefully inspected. 
In Figure~2(a) and 2(d), the peak value in 10--20 keV is $\sim$ 1.7 while $\sim$ 1.5 in 100--200 keV, 
indicating that the general trend that 
the higher photon energy is, the lower becomes the peak. 
This is neither due to incorrect background subtraction 
nor decrease in the sensitivity of the HXD, 
because the systematic uncertainty in the NXB subtraction 
is at most $\sim$ 3\% of the signal intensity even in the 100--200 keV band, 
and because we are referring to relative changes, 
instead of absolute values.  
To clarify the differences among these profiles, 
we divided the normalized shot profiles in the higher three bands by that in the 10--20 keV. 
As shown in Figure~2(e)--(g), 
the hardness ratios (relative to 10--20 keV) 
gradually decrease towards the peak, 
but suddenly return to their average values immediately after (within 0.1 s) the peak. 
Although this feature has been found in energies below $\sim$ 60 keV in Negoro et al. (1994), 
we have not only confirmed the same trend up to $\sim$ 200 keV, 
but also found that the spectral change is 
more prominent in higher energy of $E \gtrsim$ 100 keV. 

\begin{figure*}
\begin{center}
\vbox{}
\includegraphics[width=0.95\textwidth]{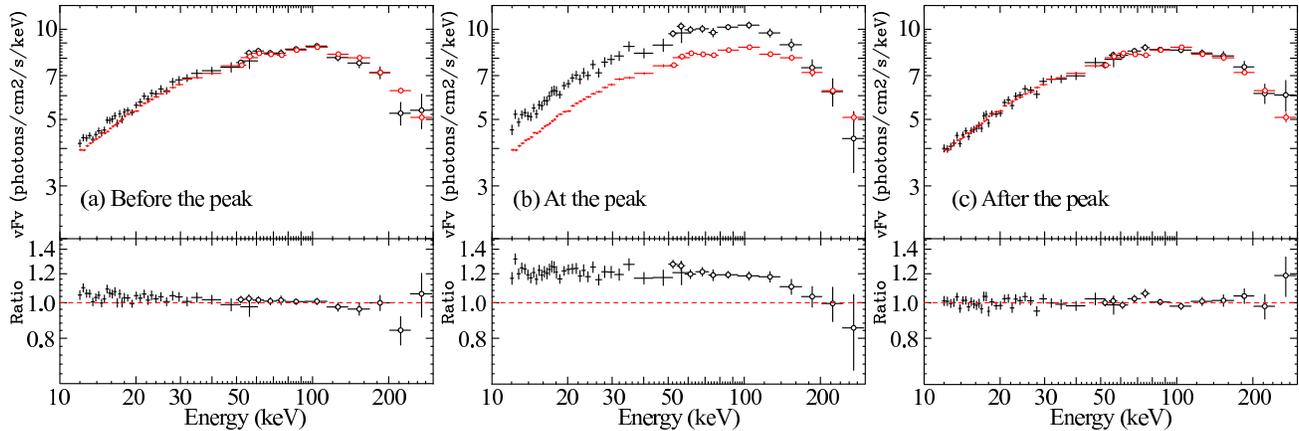}
\caption{
The background-subtracted $\nu F \nu$ spectra of the HXD, 
    accumulated over different shot phases (black). The time-averaged spectrum is given in red. 
    Panel (a), (b), and (c) show the spectra integrated from -0.25 to -0.05 s before the peak, 
    from -0.05 to 0.05 s around the peak, and 
    from 0.05 to 0.25 s after the peak, respectively. Lower panels show the ratios to the time-averaged spectrum.}
\label{fig:specfit}
\end{center}
\end{figure*}

\subsection{Quantification of the shot-phase-resolved spectra}

We then quantified its spectral change by accumulating 
the HXD spectra according to the shot phase. 
The NXB events were accumulated in the same ways and subtracted. 
Figure~3 shows three examples
of the derived shot-phase-resolved HXD spectra, corresponding to 0.15 s before, 
right on, and 0.15 s after the peak. 
The exposure at the peak is 752.4 s (7524 shots $\times$ 0.1s). 
To grasp their characteristics in a model-independent way, 
we superposed the time-averaged spectrum,
and show the ratio of the shot spectra to it in Figure~3. 
Aa shown evidently in Figure~3(b) by a clear turnover of the ratio above $\sim$ 100 keV, 
a spectral cutoff at the peak is lower than the averaged one. 
Furthermore, the spectral ratio before the peak shown in Figure~3(a) 
appears downward, while that after the peak is almost flat, 
which is consistent with the gradual softening before the peak and instant hardening at the peak 
as seen in Figure~2.
 
To consider physics underlying
this spectral evolution 
we have fitted the 13 shot-phase-resolved HXD spectra 
with a typical model of Comptonization, {\tt compps} (Poutanen \& Svensson 1996) 
in the same manner as that in Torii et al.~(2011). 
The seed photon is assumed to be a disk black body emission 
(Mitsuda et al. 1984; Makishima et al. 1986; Makishima et al.~2008) with a temperature of 0.2 keV. 
The free parameters in the fits are the electron temperature $T_{\rm{e}}$, 
the optical depth $\tau$ or the Compton $y$ parameter, 
and the normalization $N_{\rm{dbb}}$.  
Note that if $\tau$ is fixed, 
$T_{\rm{e}}$ is affected more by a spectral slope than a spectral cutoff.
To avoid such a misunderstanding, 
we kept both $\tau$ and $T_{\rm{e}}$ left free. 
As the shot-phase-resolved spectra do not have sufficient photon statistics, 
we fixed the reflection fraction $\Omega$ at the value of 0.235, 
because the obtained value from the time-averaged spectrum is $0.235_{-0.020}^{+0.021}$. 
This implies that we assumed that the reflection 
follows the primary continuum within $\sim$ 0.1 s. 
The fits to all the spectra have been successful, 
resulting in the best-fit parameters in Table~1. 
Even when considering the systematic error of the NXB in the GSO spectra, 
its contributions to the resultant values are less than $\sim$ 1\%. 

As the count rate increases on $\sim$ s time scale, 
$T_{\rm{e}}$ and $y$ decrease while $\tau$ increases; 
when the count rate starts to decrease, all the parameters appear to return to the averaged values. 
To visualize this, we plot in Figure~4 the derived parameters in Table~1, 
as well as the time-averaged ones. 
Since our composite shot profile comprises a large number of relatively small individual shots, 
the averaged parameters are close to those at $\sim$ 1 s from the peak.  
The gradual decrease in the $y$-parameter before the peak 
is consistent with the hardness decrease as seen in Figure~2. 
The decrease in $T_{\rm{e}}$ around the peak 
clearly reflect the trend that the high-energy cutoff appears lowered at the peak as seen in Figure~3(b). 
Thus, the fitting results are consistent with the hardness ratios in Figure~2 and the spectral ratios in Figure~3. 
Note that $N_{\rm{dbb}}$ also increases along the shot profile, 
though we could not confidently measure the inner radius 
without using the soft X-ray data (cf. Makishima et al.~2008).

\begin{figure}[t]
\begin{center}
\vbox{}
\includegraphics[width=0.46\textwidth]{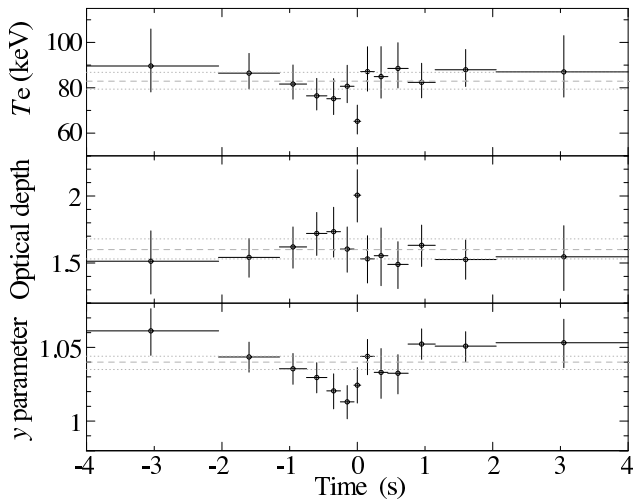}
\caption{
Time evolution of the parameters of the Comptonization, 
determined by fitting the 12--300 keV HXD spectra in the 13 shot phases 
with the Comptonization model. 
From top to bottom, the electron temperature, the optical depth, and the Compton $y$ parameter are presented.  
The 90\% range specified by the time-averaged spectrum is superposed by dotted and dashed lines.
}
\label{fig:result}
\end{center}
\end{figure}

%
%


\begin{table}[b]
\caption{The fitting results of the shot-phase-resolved spectra.}
\label{shotparatbl}
\footnotesize
\begin{center}
\begin{tabular}{rlcccc}
\hline\hline
Phase (s) &  $T_{\rm{e}}$(keV) & $\tau$  & $y$  & $N_{\rm{dbb}}$$^a$ &  $\chi^{2}_{\nu}$$^b$ \\[0.5em]                      
\hline
-3.05$\pm$1.00 & 89.6$_{-11.4}^{+16.3}$ & 1.51$_{-0.24}^{+0.23}$ & 1.061$_{-0.017}^{+0.015}$ &  5.32$_{-0.44}^{+0.56}$ & 0.92 \\
-1.60$\pm$0.45 & 86.5$_{-6.9}^{+8.7}$ & 1.54$_{-0.15}^{+0.14}$ & 1.044$_{-0.010}^{+0.010}$ &   5.78$_{-0.31}^{+0.57}$  & 1.18 \\
-0.95$\pm$0.20 & 81.7$_{-6.6}^{+8.4}$ & 1.62$_{-0.16}^{+0.15}$ & 1.036$_{-0.010}^{+0.010}$ &   5.91$_{-0.34}^{+0.59}$  & 1.12 \\
-0.60$\pm$0.15 & 76.4$_{-6.2}^{+7.7}$ & 1.72$_{-0.16}^{+0.16}$ & 1.030$_{-0.010}^{+0.010}$ &   6.01$_{-0.34}^{+0.60}$  & 1.10 \\
-0.35$\pm$0.10 & 75.2$_{-7.0}^{+8.8}$ & 1.73$_{-0.19}^{+0.18}$ & 1.021$_{-0.012}^{+0.011}$ &   6.40$_{-0.41}^{+0.63}$  & 1.20 \\
-0.15$\pm$0.10 & 80.7$_{-7.2}^{+9.2}$ & 1.60$_{-0.17}^{+0.16}$ & 1.013$_{-0.011}^{+0.011}$ &   7.64$_{-0.47}^{+0.76}$  & 1.04 \\
0.00$\pm$0.05 & 65.2$_{-5.6}^{+7.1}$ & 2.01$_{-0.20}^{+0.19}$ & 1.024$_{-0.012}^{+0.012}$  &     7.71$_{-0.51}^{+0.77}$      & 0.98 \\
0.15$\pm$0.10 & 87.1$_{-8.6}^{+10.9}$ & 1.53$_{-0.18}^{+0.17}$ & 1.044$_{-0.012}^{+0.011}$ &  7.15$_{-0.46}^{+0.71}$  & 1.09 \\
0.35$\pm$0.10 & 84.9$_{-9.5}^{+13.2}$ & 1.55$_{-0.22}^{+0.21}$ & 1.033$_{-0.018}^{+0.016}$ &  6.55$_{-0.54}^{+0.65}$ & 1.15 \\
0.60$\pm$0.15 & 88.6$_{-8.6}^{+11.3}$ & 1.49$_{-0.18}^{+0.17}$ & 1.032$_{-0.014}^{+0.013}$ &  6.45$_{-0.44}^{+0.64}$ & 1.18 \\
0.95$\pm$0.20 & 82.4$_{-6.8}^{+8.4}$ & 1.63$_{-0.16}^{+0.15}$ & 1.052$_{-0.010}^{+0.010}$ &     5.77$_{-0.35}^{+0.58}$  & 1.26 \\
1.60$\pm$0.45 & 88.0$_{-7.4}^{+8.9}$ & 1.53$_{-0.15}^{+0.14}$ & 1.051$_{-0.010}^{+0.010}$ &     5.80$_{-0.32}^{+0.58}$  & 1.22 \\
3.05$\pm$1.00 & 87.0$_{-11.1}^{+16.0}$ & 1.55$_{-0.25}^{+0.23}$ & 1.053$_{-0.017}^{+0.016}$ &    5.34$_{-0.45}^{+0.53}$ & 1.19 \\[0.5em]
\hline
All~~~~  & 82.9$_{-3.5}^{+3.9}$ & 1.60$_{-0.07}^{+0.08}$ & 1.004$_{-0.005}^{+0.004}$ &        7.04$\pm$0.18 &  1.18 \\
\hline\hline
\end{tabular}
\end{center}
\begin{itemize}
\setlength{\parskip}{0cm} 
\setlength{\itemsep}{0cm} 
\item[$^a$] In a unit of 10$^{5} R_{\rm{in}}^{2} D_{\rm{10}}^{-2} \cos\theta$, where $R_{\rm{in}}$, $D_{\rm{10}}$, and $\theta$ are the radius (km), the distance (10 kpc), and the inclination.   
\item[$^b$] The degree of freedom is 83 for the phased-sorted spectra, and 129 for the time-averaged one. 
\end{itemize}
\end{table}

\section{DISCUSSION AND SUMMARY} 

We performed the shot analysis
to extract important information on understanding rapid hard X-ray variability, 
which can not be obtained by the Fourier transform (FT) methods 
(cf. Negoro et a. 2001, Legg et al.~2012, Torii et al. 2011).
In general, FT methods are less arbitrary than the stacking analysis, 
but phase information is lost in the FT analysis.  
Further FT methods require more photons than a stacking method. 
Thus, we chose to used the stacking method  
and successfully extended the higher energy limit of the shot analysis 
up to $\sim$ 200 keV, by utilizing the HXD data as well as the P-sum mode of the XIS. 
What we found are summarized as follows: 
(1) the shot feature is found at least up to $\sim$ 200 keV with high statistical significance, 
(2) the shot profiles are approximately symmetric, 
though the hardness changes progressively more asymmetrically toward higher energies of $E \gtrsim 100$ keV, and (3) the 10--200 keV spectrum at the peak shows lower energy cutoff than the time-averaged spectrum. By quantifing this feature in terms of the single-zone 
Comptonization, we found that 
(4) as a shot develops toward the peak, $y$ and $T_{\rm{e}}$ decrease, while $\tau$ and the flux increases, and immediately past the shot peak, 
$T_{\rm{e}}$ and $\tau$ (and hence $y$) suddenly return to their time-averaged values. 


Let us consider a possible physical mechanism to explain 
the new findings and the previously-known features as well. 
The shot profile does not show any plateau at least down to $\sim$ ms (Focke et al.~2005), 
which means that most of the luminosity is released almost time-symetrically 
within $\sim$ 1 s or much shorter. 
Meanwhile, the hardness changes instantly within 0.1~s as shown in Figure~2, 
or much shorter as shown in Negoro et al.~(1994). 
These features could be explained by some physical impulse or a discrete phenomenon, 
which can change properties of the radiation source in a short time. 

When accreting matter is assumed to be an ideal and non-relativistic gas,  
the entropy of the accreting gas, $s$, with a temperature $T$ and the density $\rho$ is proportional to  $\ln(P/\rho^\gamma) = \ln (T/\rho^{\gamma-1}$),  
where $\gamma$ is the ratio of specific heat capacities  
(5/3 for monatomic gas). 
It can be interpreted that the entropy decreases in some way as the flux increases, 
but instantly increases at the peak, 
and returns to the mean value after the peak. 
This suggests 
the existence of some instant mechanism for direct entropy production (or heating). 

One of the possible ideas on the rapid intensity change 
have been considered as magnetic flares analogous to the solar corona (Galeev et al.~1979), 
and recently more sophisticated (cf., Poutanen \& Fabian~1999; Zycki~2002). 
The magnetic fields are amplified by the differential rotation of the disk, 
and rise up into the corona where they reconnect and finally liberate their energy in flare, 
making electrons accelerate. 
A typical magnetic model assumes
so-called ``avalanche magnetic flare'', in which each flare has a certain 
probability of triggering a neighboring one, 
producing long avalanches (Mineshige et al.~1994; Lyubarski, et al.~1997). 
An election temperature is expected to increase 
if magnetic reconnection accelerate electrons or protons. 
It contradicts with the gradual decrease in $T_{\rm{e}}$ before the peak, 
while agrees with the instant increase at the peak and 
an expected time scale of $\sim$ 1--100 ms 
when the energy dissipation occurs within $r \lesssim 50 R_{\rm{g}}$ ($R_{\rm{g}}$ is the gravitational radius). 
However, it is still unknown how much magnetic power is 
stored in the corona. 
Magnetic reconnection in a plunge region may be likely 
based on the global three-dimensional MHD simulation (Machida and Matsumoto~2003). 


The mass propagating model (Manmoto et al.~1996, Negoro et al.~1995) 
gives an alternative explanation, 
because the viscous time scale of the corona is $\sim$ 1 s at 100 $R_{\rm{g}}$, 
and the only $\sim$ 20\% initial perturbation 
of mass accretion rate at $\sim 100 R_{\rm{g}}$ 
can change the luminosity by $\sim$ 60\%. 
Furthermore, the mass accretion reflected as a sonic ware 
can create the latter half of the peak; 
in the theoretical view point, 
the flow passing a Bondi-type (not a disk-type) critical point 
does not always fall to the BH due to the strong outward centrifugal force 
unless angular momentum is very small (Kato, Fukue, and Mineshige~2008). 
The perturbation would start from overlapping region 
between the disk and corona, presumably $\sim$ 50--100 $R_{\rm{g}}$ (Makishima et al.~2008), 
where intense turbulence is expected due to a large gap of the pressures and temperatures. 
Since the surface density of the corona is about four orders of the magnitude smaller than that of the disk, 
a little mass transfer from the disk to the corona in the overlapping region 
could increase mass accretion into the coronae. 
Shock phenomena might be possibly working for some reason 
because there are two or more sonic points around a BH 
(Nagakura and Yamada 2009).

A rapid heating, such as magnetic reconnection, can explain the short ($\sim$ 0.1s) timescale of the shots, 
though the long ($\sim$ 1s) timescale would be related to mass accretion time scale. 
Further observational studies are needed to 
completely understand the physics causing the rapid variability. 
For instance, 
shot profiles with distinct features, such as polarization (Laurent et al.~2011) or 
a $\gamma$-ray emission (Ling et al.~1987), 
could provide a new hint, 
which would be precisely measured by new missions 
like GEMS (Black et al.~2010) and ASTRO-H SGD (Takahashi et al.~2010). 

\acknowledgements 
We would like to express our thanks to {\it Suzaku} team members, 
and also to Ryoji Matsumoto, Hiroshi Oda, and Hiroki Nagakura 
for valuable discussions. 
The research has been financed by
the Special Postdoctoral Researchers Program in RIKEN
and JSPS KAKENHI Grant Number 24740129.
S.T. and H.N are supported by Grant-in-Aid for JSPS Fellows.

\bibliographystyle{apj}

\end{document}